\documentclass[11pt]{article}
\usepackage{moriond,epsfig}

\newcommand{\EE}{\ensuremath{\mathrm{e}^+\mathrm{e}^-}}

\def\GeV{\ifmmode {\mathrm{\ Ge\kern -0.1em V}}\else
                   \textrm{~Ge\kern -0.1em V}\fi}%
\def\rts {\ensuremath{\sqrt{s}}}

\def\nnbar{\antibar{\nu}}%

\def\antibar#1{\ensuremath{#1\bar{#1}}}%

\def\a34{\cos\alpha_{34}}

\def\ee{\mathrm{e^{+}e^{-}}}

\def\EE{\mathrm{\rm \;e^+e^-\;}}

\def\xi{x_{i}}

\def\TeV{\mathrm{TeV}}

\def\O{\mathrm{\not\! O}}

\newcommand {\Be}{\begin{equation}}
\newcommand {\Ee}{\end{equation}}

\newcommand {\eqref}[1]{equation~(\ref{#1})}

\def\be{\begin{equation}}
\def\ee{\end{equation}}
\def\bea{\begin{eqnarray}}
\def\eea{\end{eqnarray}}

\begin{document}
\vspace*{4cm}
\title{Searches for Extra Dimensions  at LEP}

\author{ Marat Gataullin}

\address{Physics Department, California Institute of Technology, Pasadena,
CA 91125, USA\\E-mail: marat@caltech.edu}

\maketitle\abstracts{
Extra spatial dimensions are proposed by recent theories that postulate 
the fundamental gravitational scale to be of the same order as the
 electroweak scale.  Different final states and search strategies are used by 
LEP collaborations to search for the signs of extra spatial dimensions.  
  A brief review of the search strategies and  results from all four LEP 
experiments,   ALEPH, DELPHI,  L3 and OPAL, is given here.  No hints for 
the existence of extra dimensions are found and limits on the 
 fundamental gravitational scale are derived.  The presented results are based 
 on the data collected at LEP in 1998-2000 at centre-of-mass energies
 from $189 \GeV$\ up to $208 \GeV$. All results reported here are
 still preliminary.
}

\section{Introduction}
\subsection{Theoretical Framework}
  The Standard Model of electroweak interactions 
has been extremely successful in explaining 
the physics at its characteristic distance  $M_{ew}^{-1}$, where  $M_{ew} 
 \sim 10^{2} \GeV$  is the electroweak scale. However, this great success
has drawn attention to the hierarchy problem, {\it i.e.}, the smallness of
the electroweak
 scale as compared to  the Planck scale, $M_{Pl}
 \sim \frac{1}{\sqrt{G_N}} \sim 10^{15}\; \TeV$, which  denotes the
 characteristic scale of the gravitational  interactions. 

  It has been recently proposed~\cite{ADD}
 that this problem can be avoided  by simply removing the hierarchy.
In such models  gravity becomes strong near the  electroweak scale
and gravitational fields can propagate  in the  $n$ new
large  extra dimensions, whereas the Standard Model fields are forced
to lie on a 3-dim wall in the higher-dimensional space. Gravity thus only
appears to be weak since we can only observe its manifestations on this wall.
 The effective gravitational scale $M_D$ is then 
connected to the  Planck scale through:
\begin{equation} 
M_{Pl}^2 \sim M_{D}^{2+n} R^n,
\end{equation}
where $R$ is the size of the additional dimensions, which could be as large
 as 1~mm. Searches for  the manifestations of large extra dimensions 
at LEP are reviewed below.

\subsection{Search Strategy}

\begin{figure}
\begin{center}
  \psfig{figure=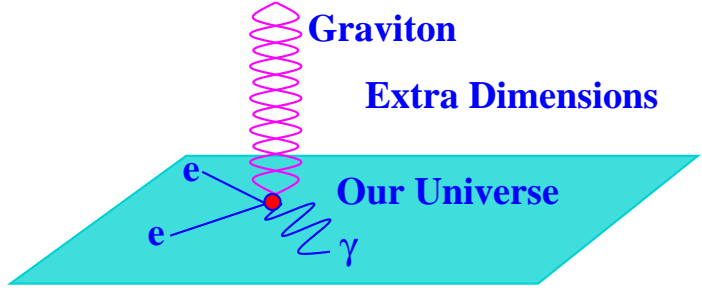,height=1.5in}
\end{center}
\caption{
Direct graviton emission diagram contributing to
 the process  $\EE \rightarrow \gamma G$.
}
   \label{fig:diag}
 
\end{figure}

  The phenomenology of large extra dimensions at LEP 
has already been studied in
 detail. Direct emission of gravitons 
leads to missing energy signatures in which a  photon or a  Z boson 
is produced
with no observable particle balancing  its transverse 
momentum~\cite{Peskin,GRW}, as shown in
Figure~\ref{fig:diag}. 

\begin{figure}
\begin{center}
  \psfig{figure=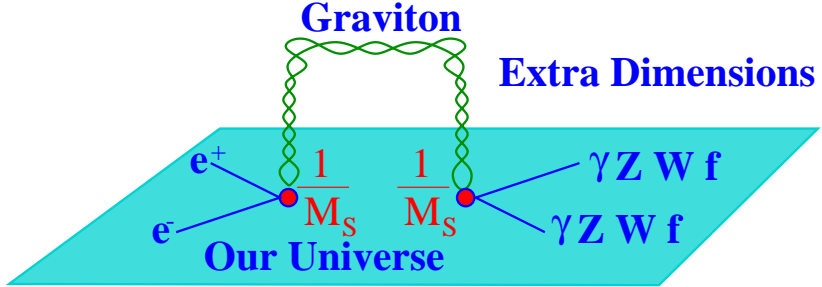,height=1.5in}
\end{center}
\caption{
Virtual  graviton exchange diagrams.
}
   \label{fig:diag1}
 
\end{figure}

The other observable effect is the anomalous production of fermion or boson
pairs in $\EE$ annihilation through $s$-channel virtual graviton exchange
as presented in Figure~\ref{fig:diag1}. The exchange of spin-2 gravitons 
modifies 
the differential cross sections  of these reactions  providing clear 
experimental signatures~\cite{GRW,Hewett,HLZ}.

\section{Direct Searches}

\begin{figure}
\begin{center}
  \psfig{figure=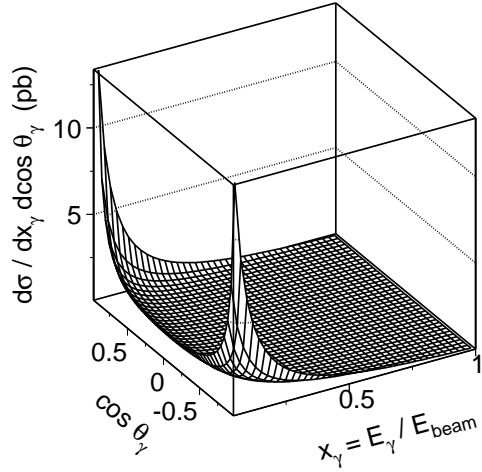,height=2.5in}
\end{center}
\caption{
Differential   cross section of the process   $\EE \rightarrow \gamma G$
     as predicted for $n=2$ and \( M_D = 1~\TeV \) at
\( \rts =      189\GeV \).
}
   \label{fig:photcross}
\end{figure}

The reaction $\EE \rightarrow \gamma G$ proceeds through $s$-channel 
photon exchange, $t$-channel  electron exchange and four-particle contact
 interaction. The differential cross section of this process depends on both 
$M_{D}$ and $n$ and is given by~\cite{GRW}:
\begin{equation}
\frac{d^2\sigma}{dx_\gamma d\cos\theta_\gamma} \; = \; \frac{\alpha}{32s}
  \, \frac{\pi^{\delta/2}}{\Gamma(\delta/2)}
 \, \left( \frac{\rts}{M_D} \right)^{\delta+2} \,
    f(x_\gamma,\cos\theta_\gamma)
 \label{eq:fxy1}
\end{equation}
with:

\begin{equation}
  f(x,y) \; = \; \frac{2(1-x)^{\frac{\delta}{2}-1}}{x(1-y^2)} \left[
    (2-x)^2 (1-x+x^2) - 3y^2x^2(1-x) - y^4x^4 \right],
  \label{eq:fxy}
\end{equation}
where $x_\gamma$ is the ratio of the photon energy to the beam energy and
$\theta_\gamma$ is
the  polar angle of the photon. The cross section increases rapidly
with smaller photon energies and angles as shown in Figure~\ref{fig:photcross}.

\begin{figure}
\begin{center}
  \psfig{figure=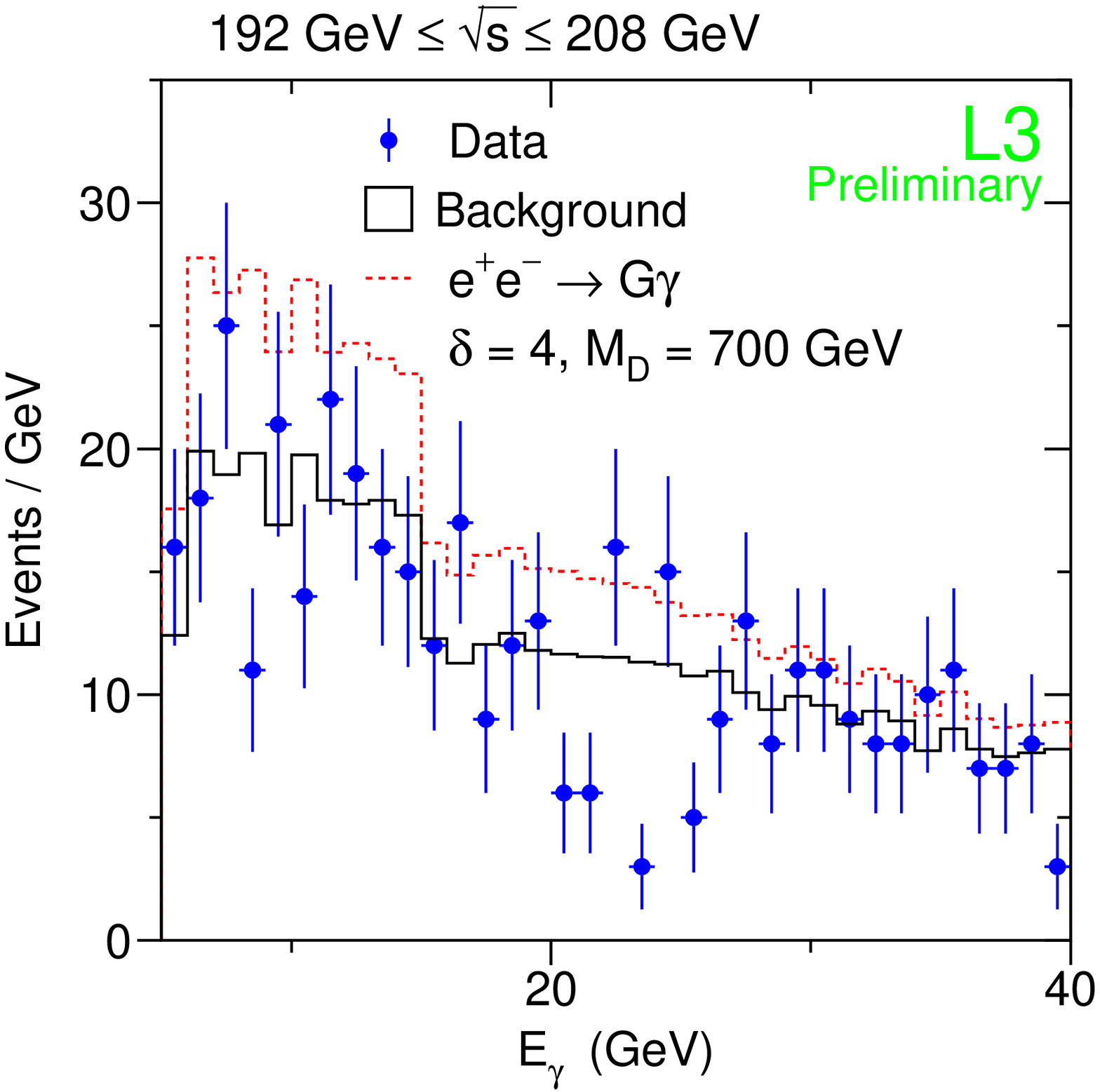,height=2.5in}\hspace{1cm}
  \psfig{figure=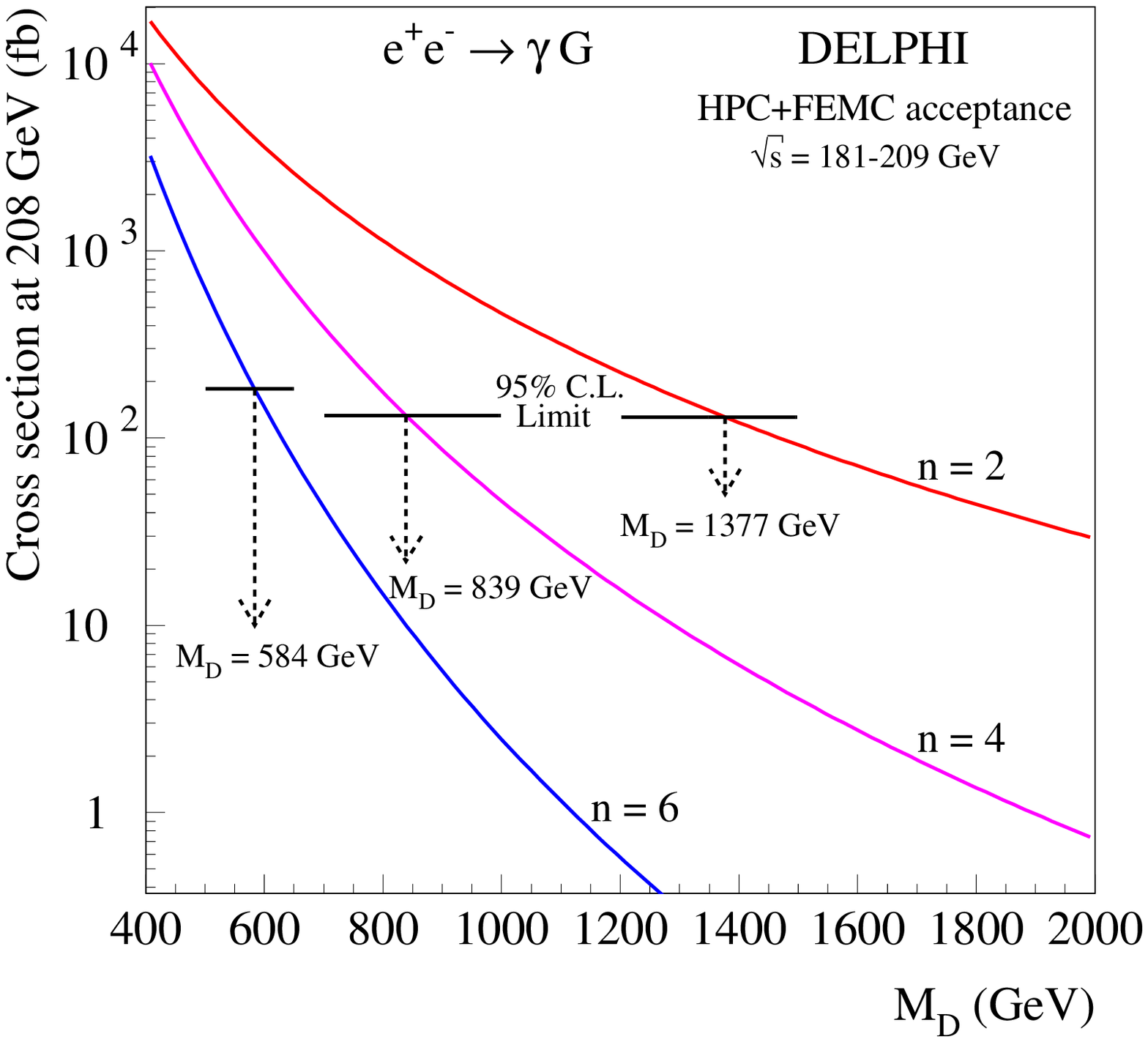,height=2.5in}
\end{center}
\caption{
  Left: L3 photon energy spectrum in real 
graviton production together with data
and expected Standard Model background;
  Right: The cross section limits at 95\% C.L. at \( \rts =      208\GeV \)
obtained by DELPHI experiment
and the expected cross sections for $n=2,4,6$ extra dimensions.
}
  \label{fig:photspec_lim}
\end{figure}

The Standard Model background for this reaction 
is dominated by the process $\EE \rightarrow \nnbar \gamma$. 
 As an  example, Figure~\ref{fig:photspec_lim} shows the expected photon
energy spectrum from real graviton production together with data
and prediction of the Standard Model background obtained by the L3 experiment.
 Limits on the parameter
$M_D$ as a function of the number of extra dimensions are then derived from 
a binned likelihood fit to the photon
energy distribution. The cross section limits on the real graviton production 
obtained by DELPHI experiment are depicted in Figure~\ref{fig:photspec_lim}.

 Lower limits~\cite{phtolim} at the  95\% C.L.
 from ALEPH, DELPHI and L3 are summarized in
Table~1~\footnote{OPAL has also
reported results using data only up  to \rts = 189~GeV.}. The expected signal
 cross section is proportional to $M_D^{-(n+2)}$ and, thus, the limits on
$M_D$ fall rapidly with the number of extra dimensions $n$.

 \begin{table}[t]

\caption{ Lower limits at 95\% C.L. on the gravitational scale $M_D$
and on the size of extra dimensions $R$, derived from direct graviton
 searches in the $\EE \rightarrow \gamma G$ channel.}
\vspace{0.4cm}
  \begin{center}
    \begin{tabular}{|c|c|c|c|c|c|c|} \hline
 \multicolumn{1}{|c|}{}       &  \multicolumn{2}{|c|}{ALEPH} &
 \multicolumn{2}{|c|}{DELPHI} & \multicolumn{2}{|c|}{L3} \\ \hline
      \multicolumn{1}{|c|}{$n$} &
      \multicolumn{1}{|c|}{$M_D$ (TeV)} & \multicolumn{1}{|c|}{$R$
      (cm)} &
      \multicolumn{1}{|c|}{$M_D$ (TeV)} & \multicolumn{1}{|c|}{$R$
      (cm)} &
      \multicolumn{1}{|c|}{$M_D$ (TeV)} & \multicolumn{1}{|c|}{$R$
      (cm)} \\ \hline
      2 & 1.28 & $2.9 \cdot  10^{-2}$   & 1.38 & $2.5 \cdot  10^{-2}$
& 1.45  &  $2.3 \cdot  10^{-2}$ \\
 \hline
          4 & 0.78 & $1.4 \cdot 10^{-9}$ & 0.84 & $1.3 \cdot 10^{-9}$ &
0.87 & $1.2 \cdot 10^{-9}$
\\ \hline
          6 & 0.57 & $5.6 \cdot 10^{-12}$ & 
0.58& $5.4 \cdot 10^{-12}$ & 0.61 &  $5.2 \cdot 10^{-12}$
\\ \hline
     
    \end{tabular}
   \end{center}
\end{table}

The direct
graviton production associated with a Z boson can also be used to
 search for the  effects of  large extra dimensions~\cite{cheung},
 complementing the search in the single photon channel. 
The L3 experiment has  searched~\cite{L3Z}  in the $\EE \rightarrow
Z G$ channel using only  hadronic Z decays selected in data collected
 in 1998 at \( \rts =      189\GeV \).   The signature of this process
is an unbalanced hadronic event with a visible mass compatible with that
of the Z. The reduced sensitivity with respect to the $\gamma G$ 
channel stems from the limited phase space available for the graviton emission
due to the mass of the Z. Limits at 95\% C.L. on the gravity scale $M_{D1}$
are given in Table~2. The gravity scale $M_{D1}$ is related but not equal to
the  $M_{D}$ parameter used in the $\gamma G$  channel, {\it
 e.g.}, in the particular
case of $n=2$ the two parameter are related as $M_{D1}^{4} = 4M_{D}^4$.

\begin{table}[t]
\caption{ Expected cross sections $\sigma_{\rm Z G}$ for the 
  Z$G$ signal $(M_{D1} = 0.5~\TeV)$, detection efficiency
    $\varepsilon$, upper limit at 
95\% C.L. $\sigma_{\rm Z G}^{\rm lim}$ on  the
 cross
    section and lower limit on the
    scale $M_{D1}$ as a function of  
    the number of extra dimensions $n$.}
\vspace{0.4cm}
  \begin{center}
    \begin{tabular}{|r|ccc|}
      \hline
      \rule{0pt}{12pt}$n$ & 2 & 3 & 4    \\ 
      \hline
      \rule{0pt}{12pt}$\sigma_{\rm Z G}$ (pb) & 0.64 & 0.081 &
      0.011  \\ 
      \hline
      \rule{0pt}{12pt} $\varepsilon$  & 0.56 & 0.56 & 0.55  \\
      \rule{0pt}{12pt}$\sigma_{\rm Z G}^{\rm lim}$ (pb) & 0.29 & 0.30 &
      0.30  \\ 
      \rule{0pt}{12pt}$M_{D1}$ (TeV) & 0.60 & 0.38 & 0.29   \\ 
      \hline
    \end{tabular}
  \end{center}
\end{table}

\section{Indirect Searches}
\subsection{Overview}
The pair production of both boson and fermions via virtual graviton exchange 
can interfere with the Standard Model production of the same final state
 particles. The  cross section can be then written as:
\begin{equation}
\left(\frac{d \sigma}{d \Omega}\right) = 
\left(\frac{d \sigma}{d \Omega}\right)_{SM} + 
 \frac{\lambda } {M_S^4}\left(\frac{d \sigma}{d \Omega}\right)_{int} 
+\frac{\lambda^2 } { M_S^8}\left(\frac{d \sigma}{d \Omega}\right)_{G},
\end{equation}
where the first term  is the Standard Model contribution, the second term
denotes the interference between the Standard Model and graviton exchange
diagrams, and the last term comes from the graviton exchange. Here $\lambda$
is a dimensionless parameter order of unity and $M_S$ is a new 
 mass scale (cut-off)
related to the  effective gravitational scale $M_D$. The exact value
of $\lambda$ depends on the details of theory. At LEP most of the experiments
 use the formalism~\cite{Hewett}, where $\lambda = +1$\ and  $\lambda = -1$\ 
cases are considered to allow for both constructive and destructive
 interference.  The dependence of  $\lambda$ on the number of extra dimensions,
$n$, is believed  to be negligible, so that the limits on  $M_S$ obtained using
this parametrization~\cite{Hewett}  are assumed to be valid for any $n$.
  All limits  are given  at 95\% C.L.
 
  Among many fermion and boson-pair final states studied at LEP,
the most sensitive channels involve Bhabha scattering $\EE \rightarrow \EE$
and the photon-pair production $\EE \rightarrow \gamma \gamma (\gamma)$.
 
\subsection{Bhabha Scattering}

\begin{figure}
\begin{center}
  \psfig{figure=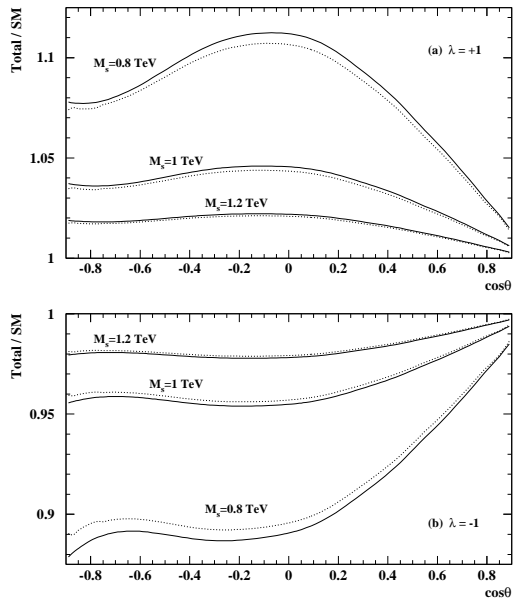,height=3.5in}
\end{center}
\caption{
Ratio of the differential   cross section of the process  
 $\EE \rightarrow \EE$ including virtual graviton exchange to the Standard
 Model expectation for (a) $\lambda = +1$\  and (b) $\lambda = -1$.
The solid curves show the Born level prediction and the dotted curves
show the prediction including radiative corrections.}
   \label{fig:opalee}
\end{figure}

The dominant diagram contributing to the reaction $\EE \rightarrow \EE$
is the $t$-channel photon exchange and the interference between this diagram
and the virtual
 graviton exchange is predicted to be large.  This combined with
the large statistics of the selected Bhabha sample at LEP makes this channel
particularly sensitive. The ratio of the differential cross section
including extra-dimensional effects to the  Standard Model expectation 
is given in Figure~\ref{fig:opalee}. The maximum deviation~\cite{opalall}
 from the Standard
 Model is observed in the central polar angle region and  reaches the value of
about 4\% for $M_S = 1~\TeV$.

\begin{figure}
\begin{center}
 \resizebox{0.4\textwidth}{7.0cm}{\includegraphics*[0cm,0cm][20.0cm,18.0cm]{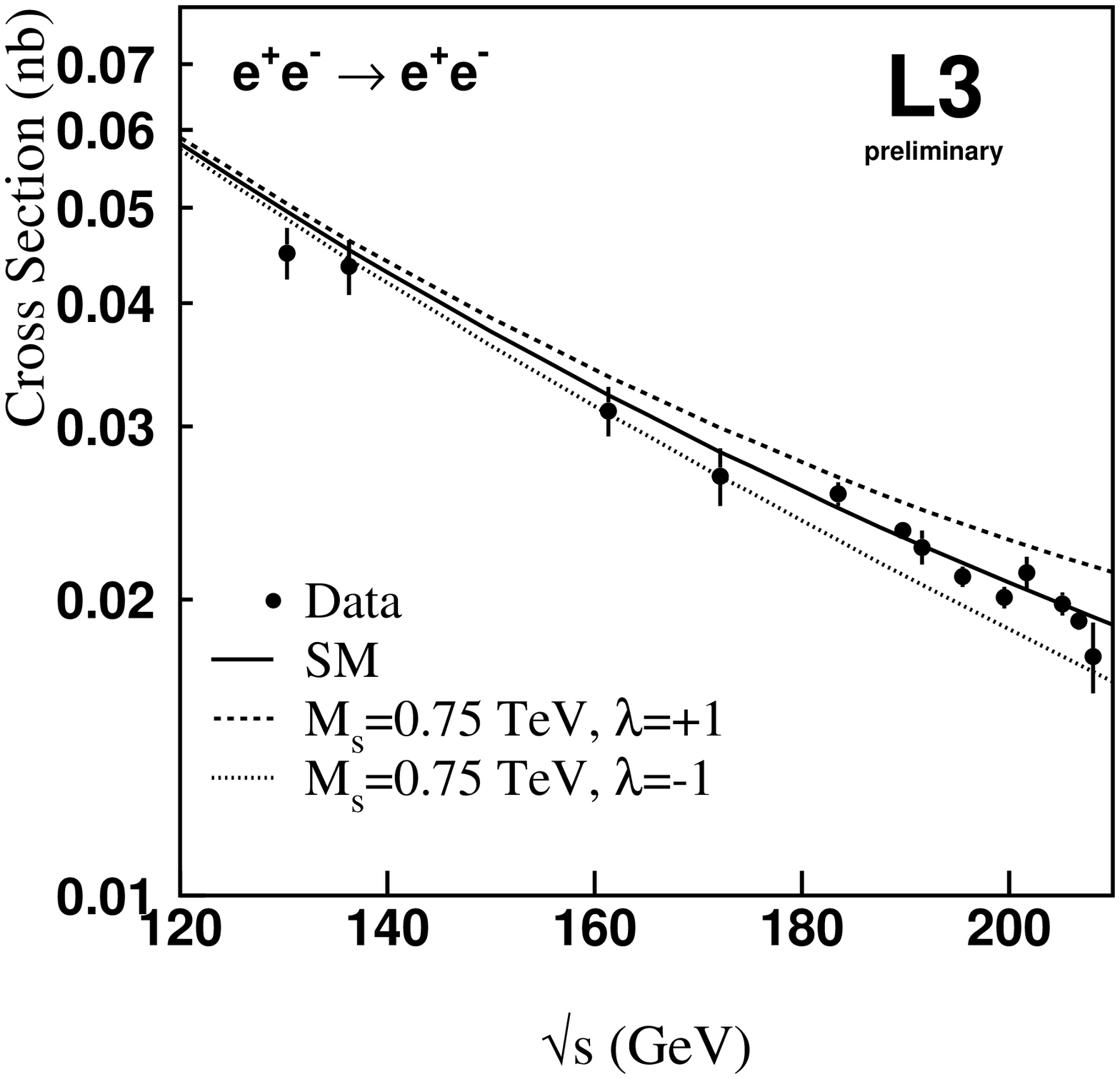}}
    \resizebox{0.4\textwidth}{7.0cm}{\includegraphics*[0cm,0cm][18.0cm,23.0cm]{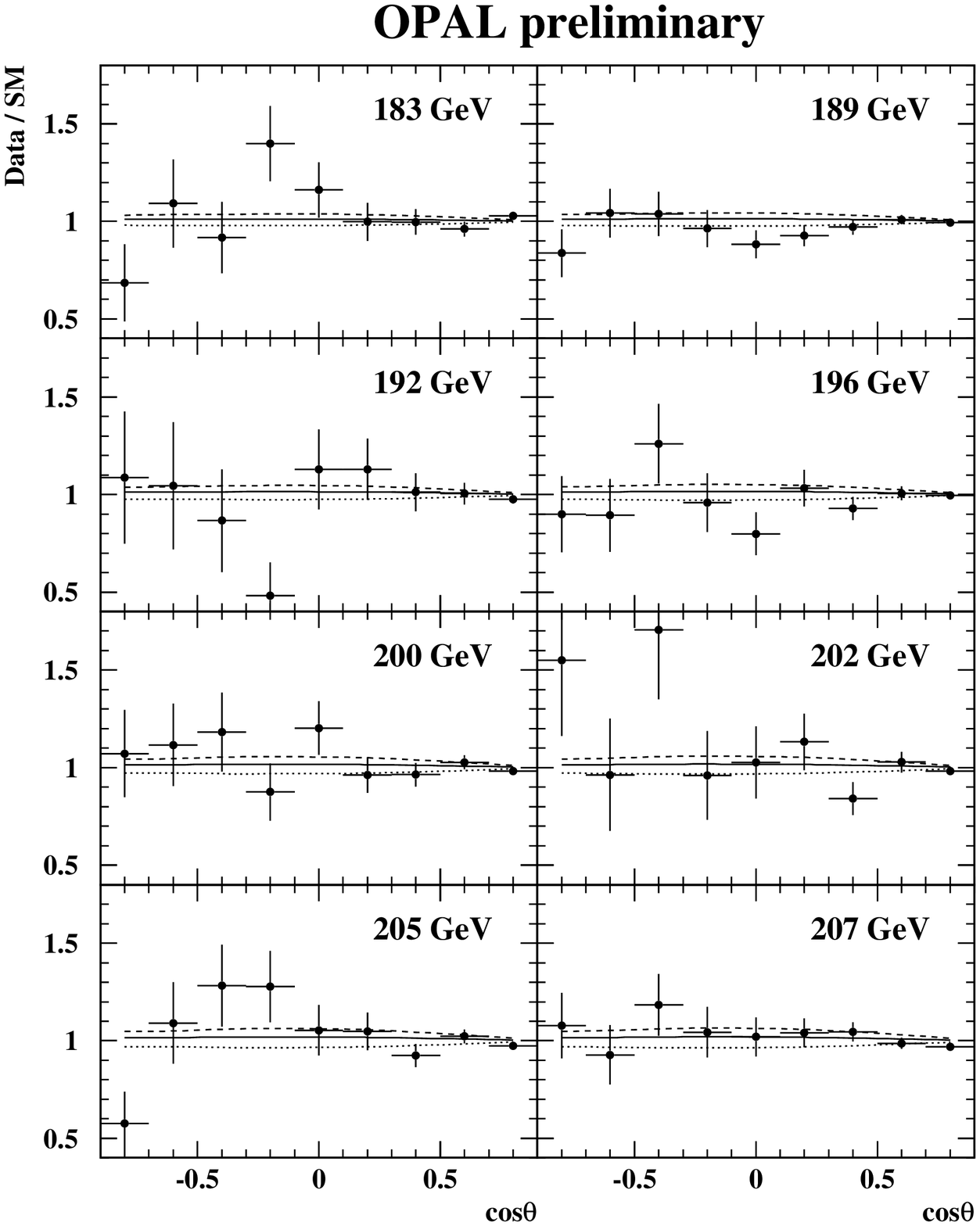}}

\end{center}
\caption{
  Left:   Bhabha cross section measured by the L3 experiment 
 as a function of $\rts$. Standard Model curve as well as the predictions
from the virtual graviton exchange are also  shown.
Right:
Ratio of the measured differential cross section to the Standard Model
 prediction obtained by OPAL at different centre-of-mass energies.
 The points show
data, whereas the dashed and dotted lines show the 95\%~C.L. limits
for $\lambda = +1$\ and  $\lambda = -1$\ respectively.
}
   \label{fig:eediff}
\end{figure}

Two LEP experiments, L3 and OPAL, have searched for the virtual graviton
exchange effects with  Bhahba samples selected using all available LEP2
 dataset~\cite{opalall,l3ee}.
 Figure~\ref{fig:eediff} shows the measured cross section distributions 
for the two experiments.
None of the two experiments see any significant deviations from the 
Standard Model expectations and  the derived limits are summarized in Table~3.
The limits on the $M_S$ are of the order of $1~\TeV$.

\begin{table}[t]
\caption{ Lower limits at 95\% C.L. on $M_S$ obtained by L3 and OPAL using
Bhabha scattering process.}
\vspace{0.4cm}
  \begin{center}
    \begin{tabular}{|c|c|c|}
      \hline
     Model &  L3 & OPAL     \\ 
      \hline 
  $ M_S(\lambda = +1)$~(TeV) & 1.06 & 1.00   \\
  \hline
$ M_S(\lambda = -1)$~(TeV) & 0.98 & 1.15 \\
 \hline
    \end{tabular}
  \end{center}
\end{table}

\subsection{Photon-Pair Production}

  The reaction $\EE \rightarrow \gamma \gamma (\gamma)$ 
proceeds via $t$- and $u$-channel QED diagrams and the cross section in 
the presence of the virtual graviton exchange is given by~\cite{AD}:
\begin{equation}
 \frac{d \sigma}{d\Omega}
= \frac{\alpha^2}{s} \frac{ \left( 1 + \cos^2 \theta \right) } { \left( 1 -\cos^2 \theta \right) } -
  \frac{\alpha \lambda s} {2 \pi M_S^4} \left( 1 + \cos^2 \theta \right)  +
  \frac{\lambda^2 s^3} {16 \pi^2 M_S^8} \left( 1 + \cos^2 \theta \right) \left( 1 - \cos^2 \theta \right). 
\end{equation}
\begin{figure}
\begin{center}
\psfig{figure=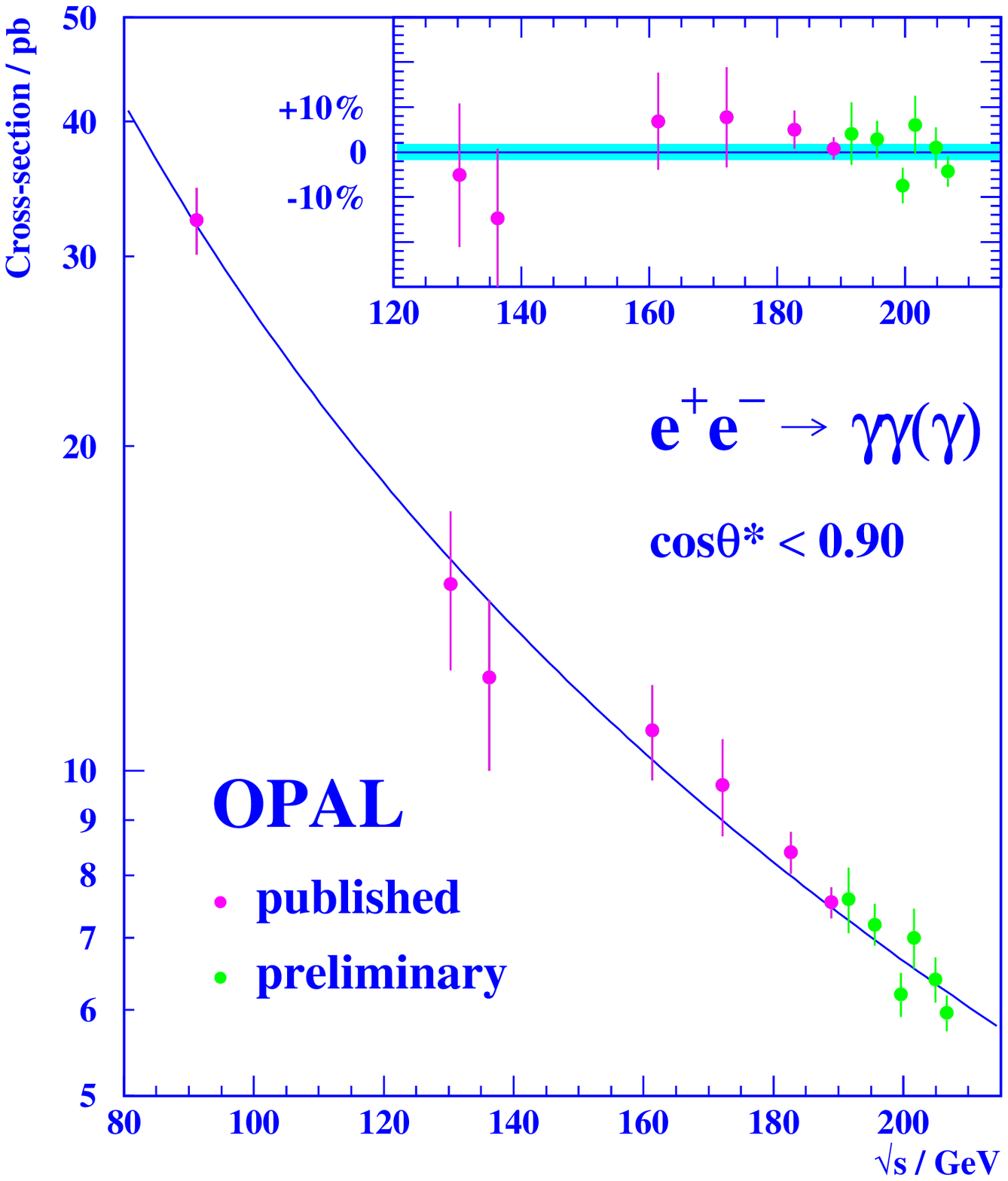,height=2.2in}\hspace{1cm}
\psfig{figure=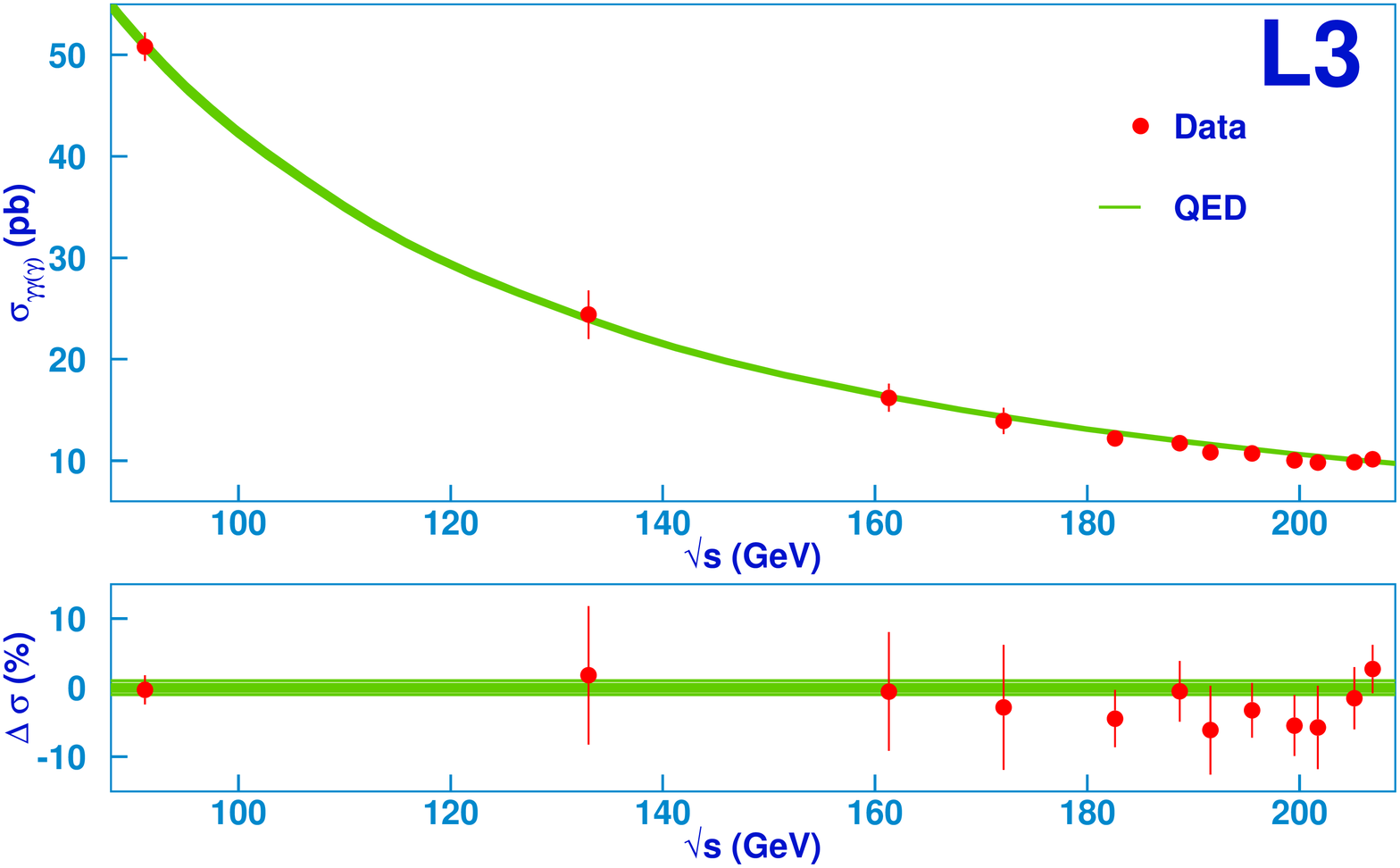,height=2.2in}
\end{center}
\caption{ Total cross section measurements for the reaction 
$\EE \rightarrow \gamma \gamma (\gamma)$ from (left) OPAL and (right) L3.
}
   \label{fig:ggcross}
\end{figure}

All four LEP collaborations have used this channel to search for the effects 
of the gravitons propagating in extra dimensions~\cite{gammall}. 
As an example, Figure~\ref{fig:ggcross} shows  the total and
differential  cross section distributions for the $\gamma\gamma$ final state
 measured by L3 and OPAL. Results from all four LEP experiments
agree very well with the Standard Model prediction. Limits on the 
cut-off scale $M_S$ are then computed using
 both the total number of the  selected events and   the distribution of the
 polar angle of the event, which is defined as 
 ${\rm cos}{\theta^{*}}=
           |{\rm sin}(\frac{{\theta}_{\gamma 1}-{\theta}_{\gamma 2}}{2})/
           {\rm sin}(\frac{{\theta}_{\gamma 1}+{\theta}_{\gamma 2}}{2})|$.

 The limits obtained by the LEP experiments
are shown in Table~4~\footnote{ALEPH  limits are based on only data collected
in 1998-99 at $\rts = 189-202 \GeV$.}.

\begin{table}[t]
\caption{ Lower limits at 95\% C.L. on $M_S$ obtained   using $\gamma\gamma$
channel.}
\vspace{0.4cm}
 \begin{center}
    \begin{tabular}{|r|c|c|c|c|}
      \hline
     Model  & ALEPH & DELPHI & L3 & OPAL    \\ 
      \hline 
  $ M_S(\lambda = +1)$~(TeV) & 0.81 & 0.82 & 0.83 & 0.83  \\
  \hline
$ M_S(\lambda = -1)$~(TeV) & 0.82 & 0.91 & 0.99 & 0.89 \\
\hline
        \end{tabular}
  \end{center}
\end{table}

\subsection{Combined Limits}
\begin{figure}
\begin{center}
  \psfig{figure=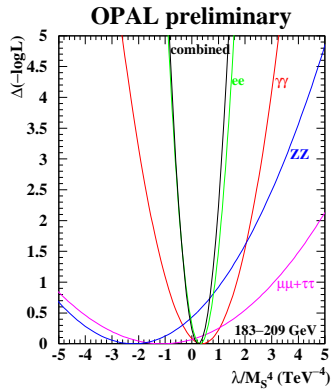,height=2.2in}
\end{center}
\caption{
Negative log likelihood curves for each channel used in the OPAL
combination and the
combined likelihood  curve. 
}
   \label{fig:opcombo}
\end{figure}

The OPAL experiment have produced combined limits~\cite{opalall} on $M_S$ 
using the following indirect search channels: $\EE$, $\gamma\gamma$ and
ZZ (using data collected at $\rts = 189-208 \GeV$), and $\mu^+\mu^-$\ and
$\tau^+\tau^-$\ (only $\rts = 189 \GeV$ data). The combined
 limit as obtained by summing  log likelihood curves from different 
channels is: $M_S(\lambda = +1) > 1.03~\TeV$ and 
$M_S(\lambda = -1) > 1.17~\TeV$.  Figure~\ref{fig:opcombo} shows that
 this limit is largely dominated by the $\EE$ channel. 

In addition, a combined limit has been obtained~\cite{bouril}
 by using the full LEP2 dataset for
the two most sensitive  $\EE \rightarrow \EE$ and 
 $\EE \rightarrow \gamma \gamma (\gamma)$ search channels.
The obtained results are summarized in Table~5. These limits are
 ``unofficial'', {\it i.e.}, they are not 
endorsed by the LEP experiments. However,
 they give  a good indication of the 
expected combined LEP sensitivity to effects of the virtual graviton exchange.

\begin{table}[t]
\caption{ ``Unofficial'' lower limits on the $M_S$ cut-off scale.}
\vspace{0.4cm}
 \begin{center}
    \begin{tabular}{|l|c|c|}
      \hline
     Process  & $ M_S(\lambda = +1)$~(TeV)  & $ M_S(\lambda = -1)$~(TeV)  \\ 
  \hline 
 Bhabha scattering & 1.13 & 1.28 \\
  \hline
QED photon pairs &   0.95 & 1.14 \\
\hline
Combined $\EE$ and   $ \gamma \gamma$ & 1.13 & 1.39 \\
\hline
        \end{tabular}
  \end{center}
\end{table}

\subsection {TeV Strings}
It has been recently shown that the additional effects due to string
 excitations
in a string theory of extra dimensions 
might compete and even overwhelm the 
effects of the virtual graviton exchange~\cite{strings}. 

The Standard Model Bhabha scattering cross section is then modified as: 
\begin{equation}
 \frac{d \sigma}{d \cos{\theta}} = \left(\frac{d 
\sigma}{d \cos{\theta}}\right)_{SM} \left|\frac{\Gamma(1-\frac{s}
{M_{Str}^2})\Gamma(1-\frac{t}{M_{Str}^2})}{\Gamma(1-\frac{s}{M_{Str}^2}-
\frac{t}{M_{Str}^2})}\right|^2, 
\end{equation}
where $M_{Str}$ is the string scale.

The L3 experiment has used cross sections and forward-backward asymmetries
from the  $\EE \rightarrow \EE$ channel   
measured using the full LEP2 dataset to derive the following limit~\cite{l3ee}
 on the
string scale:  $M_{Str} > 0.57$~TeV.

\section{Conclusion}
The four  LEP 
experiments have searched for signs of extra dimensions by looking for
 both  real graviton production and  effects of the virtual graviton exchange.
No deviations from the Standard Model are  found and preliminary 
limits in excess of 
$1~\TeV$ are set. The limits on the cut-off mass scale $M_S$ are similar to
 those obtained at TEVATRON~\cite{tevatr}, whereas the  searches
 for the 
real graviton production provide the best exclusion limits on
  new fundamental  scale of gravity $M_D$.  All  results from LEP experiments
 discussed in this paper  were taken from papers 
submitted to  the 2001 Winter Conferences. However, they are still
up-to-date, as  limits
determined using  the most sensitive channels, 
 $\nnbar \gamma$, $\EE$ and $\gamma \gamma$, 
have not been changed or finalized since~\cite{sheva}.  

\section*{Acknowledgements}
I am very grateful to the four LEP collaborations for providing
their latest results. I am indebted to S.~Mele, D.~Bourilkov, A.~Favara
and S.~Shevchenko for their invaluable help during my preparation for 
this conference.
I would like to thank the organizers of the Rencontres de Moriond, EW 2001.
 Last but not  least, I am thanking my advisor, Professor Harvey
 Newman, for his support. This work was supported in part by  the 
 U.S. Department of Energy Grant
No.~DE-FG03-92-ER40701.

\end{document}